\documentclass[
hf,
]{ceurart}

\usepackage{listings}
\usepackage{todonotes}

\usepackage{tikz}
\usetikzlibrary{arrows.meta,positioning,shapes,fit,backgrounds}
\usepackage{xspace}

\usepackage[linesnumbered,ruled,vlined]{algorithm2e}
\DontPrintSemicolon%
\usepackage{xcolor}

\newcommand{\model}{\mathcal{M}}
\newcommand{\reldom}[1]{\,\mathcal{U}(T_{#1})}
\newcommand{\reldomxj}{\reldom{x, j}}

\begin{document}

\copyrightyear{2025}
\copyrightclause{Copyright for this paper by its authors.
    Use permitted under Creative Commons License Attribution 4.0
    International (CC BY 4.0).}

\conference{SMT 2025: 23rd International Workshop on Satisfiability Modulo Theories, August 10--11, 2025, Glasgow, UK}

\title{From MBQI to Enumerative Instantiation and Back}


\author{Marek Dančo}
[%
    orcid=0009-0008-3031-113X,
]
\author{Petra Hozzová}[%
    orcid=0000-0003-0845-5811,
]
\author{Mikol\'a\v s Janota}[%
    orcid=0000-0003-3487-784X,
    url=https://people.ciirc.cvut.cz/~janotmik
]
\address[]{Czech Technical University in Prague, Prague, Czech Republic}

\begin{abstract}
    \ This work investigates the relation between model-based quantifier instantiation
(MBQI) and enumerative instantiation (EI) in Satisfiability Modulo Theories
(SMT). MBQI operates at the semantic level and guarantees to find a
counterexample to a given a model. However, it may lead to weak
instantiations. In contrast, EI strives for completeness by systematically
enumerating terms at the syntactic level. However, such terms may not be
counter-examples. Here we investigate the relation between the two techniques
and report on our initial experiments of the proposed algorithm that combines
the two.

\end{abstract}

\maketitle
\section{Introduction}

In this paper we report on work in progress in the area of quantifier instantiation for the SMT theories of (non-)linear integer
and real arithmetic combined with uninterpreted functions.\footnote{We do not consider uninterpreted sorts.}
While checking the satisfiability of formulas in these theories with quantifiers ranges from computationally expensive to undecidable,
there are many incomplete methods for quantifier instantiation that work reasonably well in practice.
It has been demonstrated repeatedly that various quantifier instantiation
techniques tend to be very complementary in performance~\cite{mbqi-sygus-tacas2025,janota2021fair}.
\emph{E-matching}~\cite{DetlefsNS05} is very powerful and often successful in
verification benchmarks but is inherently incomplete and suffers from self
capturing loops. In contrast,
\textit{enumerative instantiation}~\cite{janota2021fair,ReynoldsBF18} and
\emph{model-based quantifier
    instantiation} (MBQI)~\cite{ge-moura-cav09}, have completeness
guarantees in certain scenarios and have a semantic grounding.

In this paper we combine MBQI supported by relevant domains with enumerative instantiation.
We implemented our algorithm and compared it to cvc5~\cite{cvc5} and Z3~\cite{z3}.

\section{Preliminaries}\label{sec:preliminaries}

We assume a basic understanding of SMT~\cite{barrettSATHandbook}. In certain
special cases, such as (linear) arithmetic, there exist decision procedures for
theories with
quantifiers~\cite{Weispfenning:88,Tarski:51,Davenport:88,bjorner-lpar15,bonacina-cade2023}.
In the general case, however, quantifiers lead to
undecidability and SMT solvers rely on quantifier instantiation techniques.

We give a brief overview of MBQI.
For simplicity of presentation, assume the input formula is given in the form
\begin{equation} \label{eq:orig}
    G\land\bigwedge_j \forall\bar x^j.\,\varphi_j
\end{equation}
where $G$ is ground, and the only free variables in each $\varphi_j$ are $\bar x^j$.
Let~$\model$ be a model of~$G$ that can be expressed in the
underlying theory, and let $\varphi_j^\model$ be the evaluation of $\varphi_j$ in $\model$, meaning that
each non-variable symbol in $\varphi_j$ is replaced by its interpretation from $\model$.
We then check the satisfiability of
\[\lnot \varphi_j^\model,\]
which is quantifier-free and contains only the free
variables~$\bar x^j$ and interpreted functions and predicates.
(We note this makes the satisfiablity check easier than checking~\eqref{eq:orig}, although for some theories it is still undecidable).

If the formulas $\lnot \varphi_j^\model$ for all $j$ are \emph{unsatisfiable}, then the original
formula~\eqref{eq:orig} is true in $\model$ and hence it is satisfiable. Further, we know that $\model$ is a model
of~\eqref{eq:orig} and the algorithm terminates.
If some $\lnot \varphi_j^\model$ is
\emph{satisfiable}, then the values~$\bar v$ of $\bar x^j$ give a \emph{counterexample} to
the model~$\model$. We select ground terms $\bar t$ that have the values $\bar v$ in $\model$, instantiate $\varphi_j$ with $\bar t$,
and repeat the process with this instantiation, i.e., repeat with the
following formula:

\begin{equation*} \label{eq:step1}
    \varphi_j[\bar{x}^j\mapsto \bar{t}]\land G\land\bigwedge_j \forall\bar x^j.\,\varphi_j
\end{equation*}

\section{Algorithm}\label{sec:algorithm}

The key observation driving our work is that in arithmetic theories, MBQI has an
infinite set of theory constants, i.e. terms that denote value in a theory,
that can possibly serve as a counter-example
to the current model. Instantiating by such arbitrary constants introduces new
terms at the ground level. On the other hand, reducing the number of new terms
can have a significant positive impact on the performance~\cite{ReynoldsBF18}.
To this end, we modify MBQI to search for counterexamples by enumerating ground terms already present in the formula.
We construct a \emph{relevant domain} to guide the instantiation process.
If none of the terms in this domain serve as a counterexample, the algorithm proceeds to search for a theory constant.

Although the original MBQI paper~\cite{ge-moura-cav09} introduced an enumerative approach, MBQI implementations
in practice resort to instantiation by arbitrary theory constants; disregarding the theory constants and uninterpreted
function symbols occurring in the problem. This approach is suboptimal. For example, consider the following unsatisfiable
problem from the SMT-LIB non-incremental UFLIA benchmark set,
\[
    \forall x.\; x>0 \longrightarrow f(x) = -1000*f(x-1),\; f(0) = 1,\; f(20)<0
\]
To reach a contradiction, the solver must consider the sequence of numbers from 0 to 20. However,
both cvc5 and Z3, when running MBQI without E-matching, fail to conclude unsatisfiability.

\begin{algorithm}[th]
    \caption{MBQI with Relevant Domain}\label{alg:meta}
    \KwIn{A formula $G \land \bigwedge_j \forall\bar{x}^j.\,\varphi_j$, where $G$ is ground and each $\varphi_j$ has only the free variables $\bar{x}^j$}
    \KwOut{Satisfiability of the input formula}

    \While{true}{
        \ForEach{$j$ and each variable $x$ in $\bar{x}^j$}{
            Construct the relevant domain $\reldomxj$ \tcp*[r]{Section~\ref{sec:domain}}
        }
        Find a model $\mathcal{M}$ of $G$\;
        \If{no such model exists}{
            \Return \textbf{unsatisfiable}\;
        }
        $\texttt{is-sat} \leftarrow \texttt{true}$\;
        \ForEach{$\varphi_j[\bar{x}^j]$}{
            Fix all non-variable symbols according to $\mathcal{M}$, obtaining $\varphi_j^\mathcal{M}[\bar{x}^j]$\label{step:mod}\;
            \If{$\lnot \varphi_j^\mathcal{M}[\bar{x}^j]$ is satisfiable}{
                Using $\reldomxj$, find ground terms $\bar{t}$ s.t.\ $\lnot \varphi_j^\mathcal{M}[\bar{t}]$ is true%
                \tcp*[r]{Section~\ref{sec:instantiation}}\label{step:cex}
                $G \leftarrow G \land \varphi_j[\bar{t}]$\;
                $\texttt{is-sat} \leftarrow \texttt{false}$\;
            }
        }
        \If{\textnormal{\texttt{is-sat}}}{
            \Return \textbf{satisfiable}\;
        }
    }
\end{algorithm}

Algorithm~\ref{alg:meta} shows the general outline,  following the recipe for
MBQI (Section~\ref{sec:preliminaries}). As customary, the algorithm assumes that the
formula is split into a set of quantified subformulas ($\forall\bar
    x^j.\,\varphi_j$) and a ground part $G$, where $\varphi_j$ are quantifier-free and
contain only free variables in $\bar x^j$. Such form can be obtained by
clausification~\cite{handbookAR}, even though the formulas $\varphi_j$ do not necessarily have to be
clauses.

We note that we look for a new model of the ground part only after producing a
counterexample for all quantified formulas for which there was one. The
instantiation step and the calculation of relevant domains are explained further
in Subsections~\ref{sec:instantiation} and~\ref{sec:domain}, respectively.

We assume that constructing the formula $\varphi^{\model}_j$ is straightforward
(line~\ref{step:mod}) as we work only with arithmetic sorts and thus each
element from the model has a uniquely corresponding interpreted constant or
function. This is strictly speaking not true because of the real domain, which
is uncountable. However, in practice, solvers only produce ``well-behaved
models''.

%

\subsection{Finding a Counterexample}\label{sec:instantiation}
This section focuses on how a counterexample is chosen in Algorithm~\ref{alg:meta}
(step~\ref{step:cex}). Consider a quantified formula $\forall x^j.\varphi_j[\bar
        x^j]$ and a model $\model$. The objective is to construct a counterexample from
the relevant terms. There may be still too many of them, so we restrict those
heuristically as follows.

For each variable $x$ in $\bar x^j$, first determine a total preference ordering on
the terms in its relevant domain~$\reldomxj$ based on the current set of
formulas. The ordering is a lexicographic ordering based on how frequently does
a term appears in the formula, its depth, and in which iteration the term was
added to the formula.
We make the ordering total by appending a comparison based on the string representation of the term.
Let $\,\mathcal{P}(T_{x, j})$ denote the first third of
the terms in $\,\mathcal{U}(T_{x, j})$ with respect to this ordering. We call
these the \emph{preferred terms}.

To look for a counterexample using the preferred terms,
consider again the formula $\varphi_j^\model[\bar x^j]$, where all
non-variable uninterpreted symbols in $\varphi_j$ are interpreted by their
interpretation in the model~$\model$. Assume $\neg\varphi^\mathcal{M}_j[\bar{x}^j]$ is satisfiable.
Then, construct the following formula.
\begin{equation}\label{eq:clauses}
    \neg\varphi^\mathcal{M}_j[\bar{x}^j] \,\land \,\bigwedge_{x \,\in \,\bar{x}} \:\bigvee_{t \,\in \,\mathcal{P}(T_{x, j})} x = t^{\model}
\end{equation}

As in MBQI, the formula looks for a counterexample to the
model~$\model$, but additionally, it restricts that the counterexample is equal to
a vector of preferred terms (all this modulo what holds in the model~$\model$).
Such counterexample might not exist. Therefore, we proceed by gradually
loosening the domains of variables from~$\bar x^j$.

If the formula~\eqref{eq:clauses} is satisfiable, extract the equalities $x=t^{\model}$
that are satisfied in it and add the corresponding instantiation
$\varphi_j[\bar{x} \mapsto \bar{s}]$ to the ground part,
where each $s$ is the least term such that $s^\model = t^\model$.
In case it is unsatisfiable, one of the new clauses must be unsatisfiable.
We look for that
clause and extend the set of available terms to all terms in the relevant
domain of the corresponding variable. We do this by taking an arbitrary such clause from an unsatisfiable core
of~\eqref{eq:clauses}. This clause corresponds to some variable $x$. Next, we
permit to range over the whole set of relevant domain, not just the preferred
terms. Let $\mathcal{C} = \{x\}$ and $\mathcal{NC} =
    \bar{x}^j\setminus\mathcal{C}$. Next, construct the following formula.

\begin{equation}\label{eq:C-and-NC-equalities}
    \neg\varphi^\mathcal{M}_j[\bar{x}^j] \,\land
    \,\bigwedge_{x \,\in \,\mathcal{NC}} \:\bigvee_{t \,\in \,\mathcal{P}(T_{x, j})} x = t^{\model} \,\land
    \,\bigwedge_{x \,\in \,\mathcal{C}} \:\bigvee_{t \,\in \,\mathcal{U}(T_{x, j})} x = t^{\model}
\end{equation}

If formula~\eqref{eq:C-and-NC-equalities} is satisfiable, we have a
counterexample, as before. Otherwise, if~\eqref{eq:C-and-NC-equalities} is
unsatisfiable, continue by loosening the set of possible terms to the whole
relevant domain for variables in the core, one by one. Effectively, moving
variables from $\mathcal{NC}$ to $\mathcal{C}$
in~\eqref{eq:C-and-NC-equalities}. It this process does not yield a
counterexample, i.e.\ if the relevant domain is too restrictive, we gradually drop the
clauses altogether, effectively allowing the underlying SMT solver to pick a
theory constant (numeral). This is done analogously as before  by taking the core
of~\eqref{eq:C-and-NC-equalities} and removing variables from~$\mathcal C$, for
some $x$ whose clause appears in the core.

The approach overall can be seen a mix of enumerative
instantiation~\cite{janota2021fair} and MBQI~\cite{ge-moura-cav09}. In the first
phase, it looks for terms that are present in the current formula. If that fails
it goes back to MBQI terms. This happens per variable; therefore, the
approach is not a simple union of the two.


\subsection{Relevant Domain}\label{sec:domain}

In arithmetic benchmarks, integers/reals are often used to model the set
of some more specific objects, e.g.\ addresses on the heap.
The objective of relevant domain is to identify such scenarios.
For example, if we know that the function $f:\mathbb Z\rightarrow\mathbb Z$
operates on addresses, we want to instantiate $f(x)$ only with terms that also
represent addresses, even though $x$ ranges over all integers.

Therefore, for each quantified variable, we construct a set of candidate terms that are relevant for its instantiation.
We call such a set the \emph{relevant domain}.
To compute these sets, we use the relevant domain approach based on~\cite{ReynoldsBF18,ge-moura-cav09,KorovinSorts}.
In our approach, however, the inferred sets only serve as ``known relevant options'' for ground terms.
Our approach is not complete --- it is not always sufficient to consider only the terms in the relevant domain.
In that case we disregard the relevant domain, allowing for instantiation with an arbitrary value.

We start by creating and initializing the following domains:
\begin{itemize}
    \item $T_{x, j} = \emptyset$ for each formula $\varphi_j$ and each $x$ in $\bar x^j$, called the \emph{relevant domain} for $x$,
    \item $T_f = \emptyset$ for each uninterpreted function or predicate symbol $f$,
    \item $T_{f, i} = \emptyset$ for each uninterpreted symbol $f$ with arity $n\geq 1$ and each $1\leq i\leq n$,
    \item $T_u = \{u\}$ for each ground term $u$ occurring anywhere in the input formula.
\end{itemize}

Then we follow the rules shown below to merge these sets.
We represent the merged set by a union containing the sets it was formed from, and we abuse to notation to refer to the union by any of its constituting sets.
For example, the result of merging $T_f$ and $T_{x, 1}$ is $\{T_f, T_{x,1}\}$, and the result of a subsequent merge of $T_f$ and $T_a$ is $\{T_f, T_{x,1}, T_a\}$.
To denote that $T_1$ and $T_2$ should be merged, we write $\texttt{merge}(T_1, T_2)$.

Although the sets $T_{x, j}$ are initially empty, after applying the merging rules exhaustively,
they may become members of a union containing some sets of ground terms $T_t$.
Let us define $\reldomxj$ as $\{t ~|~ t \text{ is a ground term}\land \exists U.\ T_t\in U \land T_{x,j}\in U\}$ -- i.e., the set of ground terms $t$ whose corresponding set $T_t$ belongs to the same union as $T_{x,j}$. 
We consider all the terms of $\reldomxj$ to be relevant for instantiating the
quantified variable $x$ in the subformula $\varphi_j$.
If $\,\mathcal{U}(T_{x, j})$ is empty, we add a default theory constant to it; in our case of arithmetic, this default is 0.

We now explain the merging rules.
First, we define the function $\mathit{tls}$ that assigns to each term\footnote{In practice, SMT-LIB benchmarks typically do not include division operations, so we disregard such terms.}
its \emph{top level symbol}:

\[
    \textit{tls}(t) =
    \begin{cases}
        u                 & \text{if } t \text{ is a ground term } u                                                                   \\
        x^j               & \text{if } t \text{ is a quantified variable } x \text{ in subformula $j$}                                 \\
        g                 & \text{if } t \text{ is } g(t_1,\dots,t_n) \text{, where } g \text{ is uninterpreted and $t$ is not ground} \\
        \textit{tls}(t_1) & \text{if } t \text{ is } t_1 * \dots * t_n \text{ and $t$ is not ground}                                   \\
        \textit{tls}(t_1) & \text{if } t \text{ is } t_1 + \dots + t_n \text{ and $t$ is not ground}                                   \\
        \textit{tls}(t_1) & \text{if } t \text{ is } t_1 - t_2 \text{ and $t$ is not ground}                                           \\
    \end{cases}
\]
Next, we define the function $T^*$ that maps terms to their respective sets based on their top level symbols.
\[
    T^*(t) =
    \begin{cases}
        T_u      & \text{if } \textit{tls}(t)  \text{ is a ground term } u                                                                     \\
        T_{x, j} & \text{if } \textit{tls}(t) \text{ is a quantified variable } x^j \text{ in subformula } j                                   \\
        T_f      & \text{if } \textit{tls}(t)  \text{ is an uninterpreted function (predicate) symbol } f \text{ and } t \text{ is not ground} \\
    \end{cases}
\]

The rules merge the introduced sets based on where a term appears within the formula. Specifically, we
distinguish between terms that occur as arguments of uninterpreted function symbols, arithmetic operations,
or relational operators such as $\{=, \,\leq, \,<\}$.
We apply these rules based on term occurrences in the whole formula $G\land\bigwedge_j \forall\bar x^j.\,\varphi_j$.

\begin{center}\setlength{\tabcolsep}{1.5em}
    \begin{tabular}{ll}
        Term occurring in the formula & Operation(s)                                                 \\ \hline
        $f(t_1,\dots,t_n)$            & \texttt{merge}$(T_{f, i}, T^*(t_i))$ \ for \ $1\leq i\leq n$ \\
        $t_1 + \dots + t_n$           & \texttt{merge}$(T^*(t_1), T^*(t_i))$ \ for \ $2\leq i\leq n$ \\
        $t_1 - t_2$                   & \texttt{merge}$(T^*(t_1), T^*(t_2))$                         \\
        $t_1 * \dots * t_n$           & \texttt{merge}$(T^*(t_1), T^*(t_i))$ \ for \ $2\leq i\leq n$ \\
        $t_1 = t_2$                   & \texttt{merge}$(T^*(t_1), T^*(t_2))$                         \\
        $t_1 \leq t_2$                & \texttt{merge}$(T^*(t_1), T^*(t_2))$                         \\
        $t_1 < t_2$                   & \texttt{merge}$(T^*(t_1), T^*(t_2))$                         \\
    \end{tabular}
\end{center}

%
%

As an example, consider the following SMT formula.
\[
    \underbrace{(f(3, a) \geq 4 + g(b))}_{G} \,\land \,\underbrace{(\forall x, y. \,f(x, y) < x + g(y))}_{\forall x, y.\varphi_1} \,\land \,\underbrace{(\forall x. \,g(x) = g(x + 2))}_{\forall x.\varphi_2}
\]
Because $x$ and $g(y)$ are summed in $\varphi_1$, we apply the operation $\texttt{merge}(T_{x, 1}, T_g)$. Similarly, because $x$ is also the first argument of $f$, we apply
$\texttt{merge}(T_{f, 1}, T_{x, 1})$. Using these two rules we get the union $\{T_{x, 1}, T_{f, 1}, T_g\}$.
After performing all the possible merges, we  get two disjoint unions of sets.
\begin{center}
    $\{T_{x, 1}, T_{f, 1}, T_g, T_f, T_4, T_3, T_{(4 + g(b))}\}$ \\
    $\{T_{y, 1}, T_{f, 2}, T_{g, 1}, T_a, T_b, T_{x, 2}, T_2\}$  \\
\end{center}
Since the first set contains both $T_{x,1}$ as well as $T_4, T_3$, and $T_{(4 + g(b))}$, we get that
$\,\mathcal{U}(T_{x, 1}) = \{3, \,4, \,\mbox{$4 + g(b)$}\}$, and similarly $\,\mathcal{U}(T_{y, 1}) = \,\mathcal{U}(T_{x, 2}) = \{2, \,a, \,b\}$.

Note that we repeat the merging after each instantiation step (line~12 of Algorithm~\ref{alg:meta}).
This is because an instantiation may give us more information about the relevant domains.
For example, instantiating $x = 2$ in the last subformula yields a ground formula
$g(2) = g(4)$. After adding it to the ground part, we apply $\texttt{merge}(T_{g,1}, T_4)$ and merge the two unions of sets into one.

\section{Preliminary Experiments}\label{sec:experiments}
We initially implemented the algorithm described in Section~\ref{sec:algorithm} as a standalone Python module using PySMT~\cite{pysmt2015},
with Z3~\cite{z3} as its backend SMT solver for quantifier-free queries.

\paragraph{Benchmarks.}
For preliminary evaluation, we used a subset of benchmarks from~\cite{imo-cade2025} based on problems from the International Mathematical Olympiad, encoded in SMT-LIB~\cite{smt2} using UFNRA logic.
We chose 22 problems from this set for which we manually confirmed they are unsatisfiable.

\paragraph{Experimental setup.}
We compared the performance of our implementation with the SMT solvers cvc5~\cite{cvc5}, Z3~\cite{z3}, and Interpol~\cite{smtinterpol}, and with the first-order theorem prover Vampire~\cite{vampire}.
For cvc5, we used 5 instances with the following configurations, later denoted as cvc5-\{1,2,3,4,5\}:
\begin{enumerate}
    \item default mode
    \item \texttt{-{}-enum-inst}
    \item \texttt{-{}-mbqi}
    \item \texttt{-{}-no-e-matching -{}-enum-inst}
    \item \texttt{-{}-simplification=none -{}-enum-inst}
\end{enumerate}
We ran all the other solvers in default mode.
The experiments were performed on machines with two AMD EPYC 7513 32-Core
processors and with 514~GiB~RAM.
We used the time limit of 600 seconds per problem for each solver.

\begin{table}[t]
    \caption{Numbers of solved problems out of total 22.}\label{tab:results}
    \begin{tabular}{ccccccccc}
        This work & cvc5-1 & cvc5-2 & cvc5-3 & cvc5-4 & cvc5-5 & Interpol & Vampire & Z3 \\ \hline
        9         & 5      & 10     & 6      & 13     & 10     & 5        & 15      & 10
    \end{tabular}
\end{table}

\paragraph{Results.}
We display the numbers of solved problems in Table~\ref{tab:results}.
Overall, Vampire solves most problems.
Our prototype is more-or-less competitive with the SMT solvers, solving more problems than cvc5-1, cvc5-3, and Interpol, one less problem than cvc5-2, cvc5-5, and Z3, and four problems less than cvc5-4.
Interestingly, our prototype also solves one problem which none of the other SMT solvers solve -- the only other tool to solve it is Vampire.
We therefore conclude that our preliminary results show promise and we plan to further develop our approach.

\section{Conclusion and Future Work}\label{sec:conclusion}
In this paper we gave preliminary results on MBQI combined with a custom
relevant domain approach. The approach resembles enumerative
instantiation~\cite{ReynoldsBF18,janota2021fair} and MBQI~\cite{ge-moura-cav09},
as it focuses on the ground terms currently present in the formula, and only
abandons those if no counterexample (meaning a good instantiation) exists.
The approach can be naturally generalized to other non-arithmetic sorts.

The current prototype shows promise since it allows solving small but difficult
problems. We plan to investigate improvements to the implementation that would
enable solving also larger problem instances and integrating it into an
existing solver.

\section*{Acknowledgements}
This work is supported by the Czech MEYS under the ERC~CZ project no.~LL1902
\emph{POSTMAN} and by the European Union under the project \emph{ROBOPROX}
(reg.~no.~CZ.02.01.01/00/22\_008/0004590) and by the \emph{RICAIP} project
that has received funding from the European Union's Horizon~2020 research and
innovation programme under grant agreement No~857306.

\section*{Declaration on Generative AI}
The authors used ChatGPT and Grammarly for grammar and spell-checking. After using
these tools, the authors reviewed and edited content as needed and take full
responsibility for the publication's content.

\bibliography{refs}
\end{document}